\documentclass[aps,twocolumn,showpacs]{revtex4}
\usepackage{graphicx}

\begin{document}
\title{Tunable THz-whistle built on two-dimensional Helmholtz resonator}
\author{M.V. Cheremisin}
\affiliation{A.F.Ioffe Physical-Technical Institute, St.Petersburg, Russia}
\date{\today}
\begin{abstract}
We propose the turnable THz generator built on two-dimensional analog of conventional Helmholtz resonator. The generator excitation is provided
by current flow similar to whistle in conventional hydrodynamics. The output frequency, hydrodynamic parameters and practical realization of
THz-whistle are discussed regarding the possible applications.
\end{abstract}
\maketitle

\section{Introduction}
\label{Introduction}
Plasma oscillations in 2D electron gas were predicted in the 60th by F.Stern\cite{Stern67} for ungated and, then analyzed for
gated systems\cite{Chaplik72}. Over the next decade the plasmon assisted infrared absorbtion\cite{Grimes76,Allen77,Theis78,Heitmann82}
and emission\cite{Tsui80} has been reported. Since Stern's pioneering discovery the enormous efforts were done to clarify the plasmon behavior found to be influenced by magnetic field\cite{Chiu74,Nakayama74,Volkov88}, retardation effects\cite{Kukushkin03}, geometry\cite{Kosevich88,Fetter86,Zabolotnykh22} and quality\cite{Falko89,Cheremisin17}.

The enhanced interest to plasma oscillations in 2D systems is triggered by apparent simplicity for creation of tunable generator(detector)\cite{Dyakonov93,Petrov20} in terahertz frequencies range. For this purpose, the 2D system with carrier density, $n$, controlled by the gate-to-channel voltage $U_{g}$ is usually addressed. Neglecting retardation effects\cite{Chaplik15}, the plasmon dispersion is linear\cite{Chaplik72,Burke00}:
\begin{equation}
\omega_{p}=vk,
\label{plasmon}\\
\end{equation}
where $v=\sqrt{\frac{4\pi n e^{2}d}{m\varepsilon}}$ and $k$ is the plasma wave velocity and wave vector respectively, $m$ is the effective mass of electrons. Then, $n=C_{g}(U_{g}-U_{T})/e$ is the 2D carrier density, $U_{T}$ is the threshold voltage\cite{Shur89}, $C_{g}=\frac{\varepsilon}{4\pi d}$ is the capacitance per unit area, $d$ is the width and $\varepsilon$ is the dielectric permittivity of gate-to-channel spacing. Equation(\ref{plasmon}) implies the exact definition of plasmon wave vector which depends on sample geometry being reciprocal to typical scale of the in-plane sizes.

Usually, the outward source of electromagnetic field and, moreover, the metal gratings\cite{Tsui80} are used to excite 2D plasmon. Even recent studies\cite{Kukushkin20} consist of electromagnetic radiation directed by either waveguide or coupled wire antenna. Alternatively, a current driven mechanism for plasmon excitation was proposed\cite{Dyakonov93} for submicron 2D sample lengths but, then was revised\cite{Dyakonov08} because of actual sample widths are three orders of magnitude more. A broadband turnable THz-source is not realized yet. Hereafter, we intend to avoid these complications by making use of Helmholtz resonator\cite{Helmholtz1885}.

\begin{figure}[tbp]
\begin{center}\leavevmode
\includegraphics[width=1.0 \linewidth]{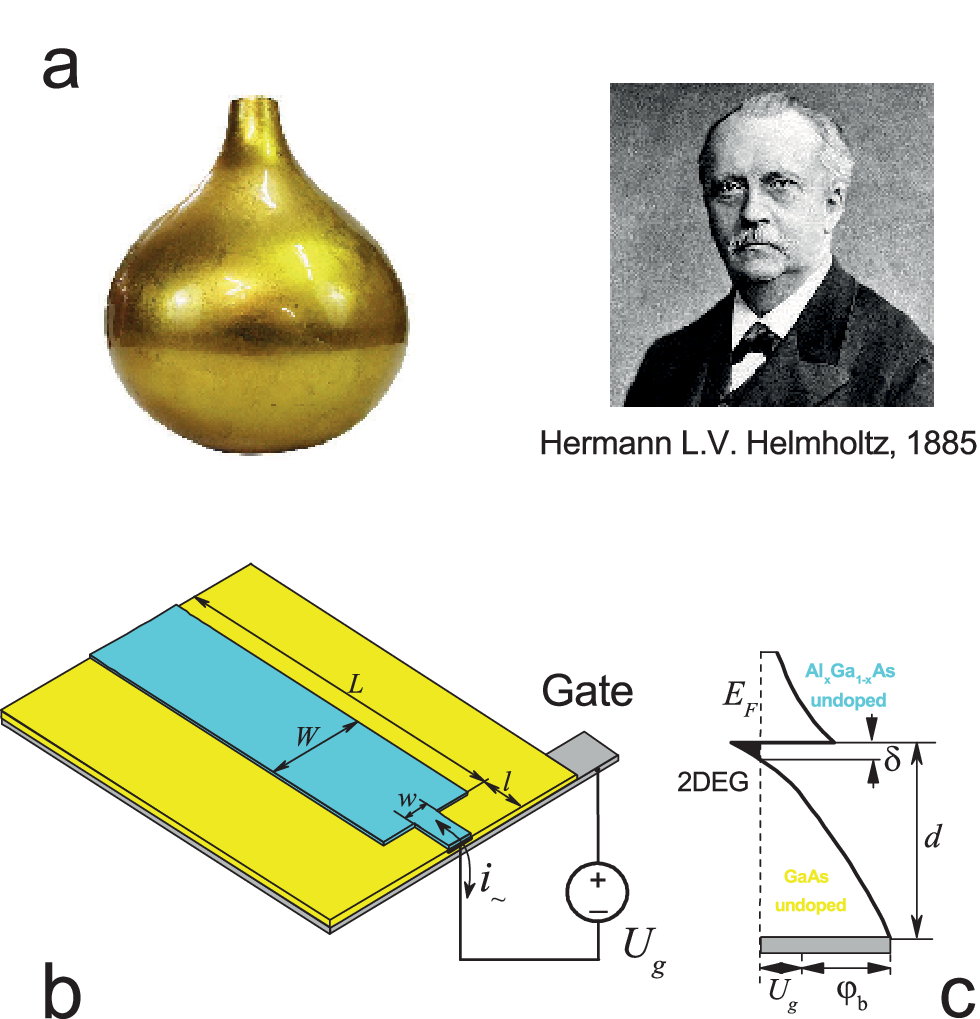} \caption[]{\label{Fig1} a.Acoustic Helmholtz resonator; b. Two-dimensional analog; c. Band diagram of AlGaAs/GaAs inverse\cite{Shur89} heterostructure with backside metallic gate, $E_{F}$ and $\phi_{b}$ are the Fermi and the band energy respectively.}
\end{center}
\end{figure}

\section{Classical Helmholtz resonator}
\label{Classical Helmholtz resonator}
The classical Helmholtz resonator consists of a rigid-walled cavity of volume $V$, with narrow neck of area, $s$, and length, $l$ shown in Fig.\ref{Fig1},a. The device operation is \emph{only} possible when the compressible gas mass transfer occurs between the resonator and the environment. The operation of the device can be understood\cite{Raichel2006} by considering a harmonic oscillator with a single degree of freedom. The air mass in the neck $m=\rho sl$ serves as moving piston while the large volume of air in a cavity plays the role of spring rigidity $k=\rho s^{2}c^{2}/V$, where $\rho$ is the volumetric density, $c$ this is the speed of sound in the air. Following the analogy with mechanical spring oscillator the resonance frequency $\omega_{3H}=(k/m)^{1/2}$ yields
\begin{equation}
\omega_{3H}=c\sqrt{\frac{s}{Vl}}.
\label{3H_Helmholtz}\\
\end{equation}
The frequency depends on the cavity volume, not the shape. For rectangular neck $l \simeq \sqrt{s}$ the frequency of 3D Helmholtz
resonator $\omega_{3H} \sim \frac{c}{V^{1/3}}\sqrt{\frac{l}{V^{1/3}}}$ is determined by average cavity size $V^{1/3}$ multiplied by a small factor. The wavelength $\lambda=2\pi c/\omega_{3H}$ of resonator excitations is larger than typical device size. Undoubtedly, there are additional overtones caused by standing waves in cavity being, however, not harmonics of the fundamental one specified by Eq.(\ref{3H_Helmholtz}). Importantly, for realistic acoustic resonator the neck length $l$ embedded into Eq.(\ref{3H_Helmholtz}) can be modified as $l+\Delta l$, where $\Delta l \ll l$ is so-called "end correction" term\cite{Fletcher1991} caused by sound excitation of finite size $\Delta l \sim \sqrt{s}$ of the flanged pipe. Therefore, Eq.(\ref{3H_Helmholtz}) remains valid as well for resonator with cropped neck, i.e. when $l=0$.

The conventional Helmholtz resonator demonstrates unprecedented quality with sole losses caused by sound irradiated into surrounding medium, not the gas-to-wall friction and(or) viscosity. According to Ref.\cite{Raichel2006} the quality factor $Q=\lambda l/s \gg 1$ depends on device sizes. Helmholtz used by himself a set of calibrated resonators acting as high quality sound spectrometer. In musical practice, a hollow soundboard of string instrument(guitar, lute, violin etc.) with a hole(s) coupled to the outside air comprises the Helmholtz resonator. Moreover, even a human whistle\cite{Wilson1970} operates an accordance to Eq.(\ref{3H_Helmholtz}). In later case the gas flux in whistle interior becomes unstable\cite{Wilson1970} under vortex formation when the jet velocity exceeds a certain threshold. The reason is that the Fourier expansion of an arbitrary vortex flow represents a continuum spectra, including harmonics coinciding with Helmholtz resonator. Unlike other sound sources the human whistle excludes sharp edges or vibrating elements. Therefore, whistle mechanism may provide a reliable way to generate monochromatic tone compare to well known edge pulsating jet and reed excitation\cite{Fletcher1991}.

\section{Two-dimentional Helmholtz resonator}
\label{Two-dimentional Helmholtz resonator}

Starting from Gurzhi paper\cite{Gurzhi63} the essential theoretical\cite{Fetter86,Dyakonov93,Alekseev16} and experimental efforts\cite{Molenkamp95,Bandurin16,Sulpizio19,Steinberg22} have been made to reveal hydrodynamic behavior of 2D electrons. Recent experiments\cite{Steinberg22} on vortices visualization and transition from ohmic to hydrodynamic flow support the carrier fluidity of high quality 2D systems. It is generally accepted that the electron fluidity appears when the length of electron-electron interactions, $l_{ee}$, is less than typical sample size and, moreover, much less than the length, $l_{p}$, responsible for momentum transfer from electrons to lattice. The hydrodynamics approach implies that Navier-Stokes equation
\begin{equation}
\frac{\partial {\bf V}}{\partial t}+({\bf V}\nabla){\bf V}=\eta\triangle{\bf V}-\frac{{\bf V}}{\tau}+\frac{e}{m}{\bf E}
\label{Navier-Stokes}\\
\end{equation}
is valid. Here, $\bf{V}$ is the flux velocity, $\triangle$ the Laplace operator and $\bf{E}$ the in-plane electric field. Then, $\eta=V_{F}l_{ee}/4$ is the kinematic viscosity\cite{Alekseev16}, $V_{F}$ is the Fermi velocity and, finally $\tau=l_{p}/V_{F}$ denotes the carriers momentum relaxation time. For steady state flow $\frac{\partial}{\partial t}=0$ one may estimate the all terms embedded into Eq.(\ref{Navier-Stokes}). Importantly, the sample size plays the crucial role defining the length scale of the problem. For simplicity, we will consider a rectangular sample $W \times W$ in two-dimensional plane. Introducing dimensionless velocity $u=V/V_{F}$ one may roughly estimate the magnitude of Euler, viscosity and friction terms embedded into Eq.(\ref{Navier-Stokes}) as it follows
\begin{equation}
u^{2}, \qquad \frac{l_{ee}}{4W}u, \qquad \frac{W}{l_{p}}u.
\label{Terms_Ratio}\\
\end{equation}

\begin{figure}[tbp]
\begin{center} \leavevmode
\includegraphics[width=1.0 \linewidth]{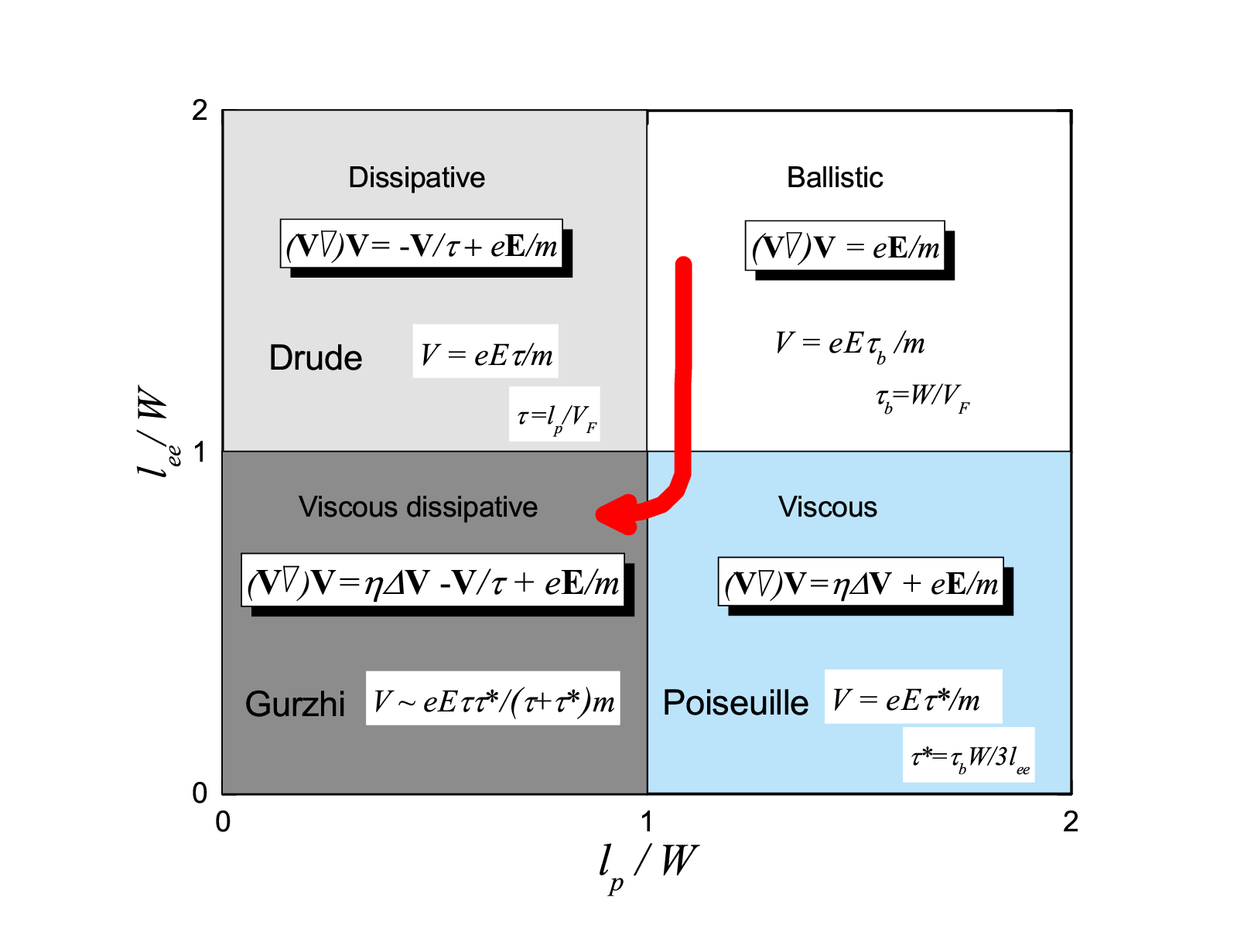} \caption[]{\label{Fig2} Diagram of 2D hydrodynamic flow mode. Arrow demonstrates schematically a temperature dependent change of transport mode expected for fixed sample size depicted in Fig.\ref{Fig3}.}
\end{center}
\end{figure}

Predominance of a certain term(s) in Eq.(\ref{Terms_Ratio}) defines a type of hydrodynamic flow driven by applied force $\frac{eEW}{mV_{F}}$.
More specifically, the ratios $\frac{\l_{ee}}{W},\frac{\l_{p}}{W}$ are crucial. Evidence shows that the electrons moved without collisions on scales exceeding the sample length, i.e when $l_{p},l_{ee}>W$ can be viewed as ballistic. Then, if the scattering length of either $l_{p}$ or $l_{ee}$ exceeds the sample size $W$ the friction(viscosity) term can be dropped in Eq.(\ref{Navier-Stokes}) leading to viscous( dissipative ) regimes respectively seen in Fig.\ref{Fig2}. Finally, for frequent collisions $l_{p},l_{ee}<W$ one recovers the combined regime which includes both viscous and dissipative terms.

Let us turn to possible nonlinear effects governed by velocity $u$ strength. Note that Reynolds number $R=u\frac{4W}{l_{ee}}$ given by Euler to viscosity term ratio in Eq.(\ref{Terms_Ratio}) is used to account for nonlinear in velocity effects in conventional hydrodynamics. By contrast, one may introduce the ratio $C=u\frac{l_{p}}{W}$ of the Euler to friction term being a measure nonlinear effects within dissipative regime. When either $R \ll 1$ or $C \ll 1$ condition is fulfilled one recovers Poiseuille and Drude flow respectively. For combined set $R,C\ll 1$ one obtains the so-called Gurzhi flow mode depicted in Fig.\ref{Fig2}.

\begin{figure}[tbp]
\begin{center}\leavevmode
\includegraphics[width=1.0\linewidth]{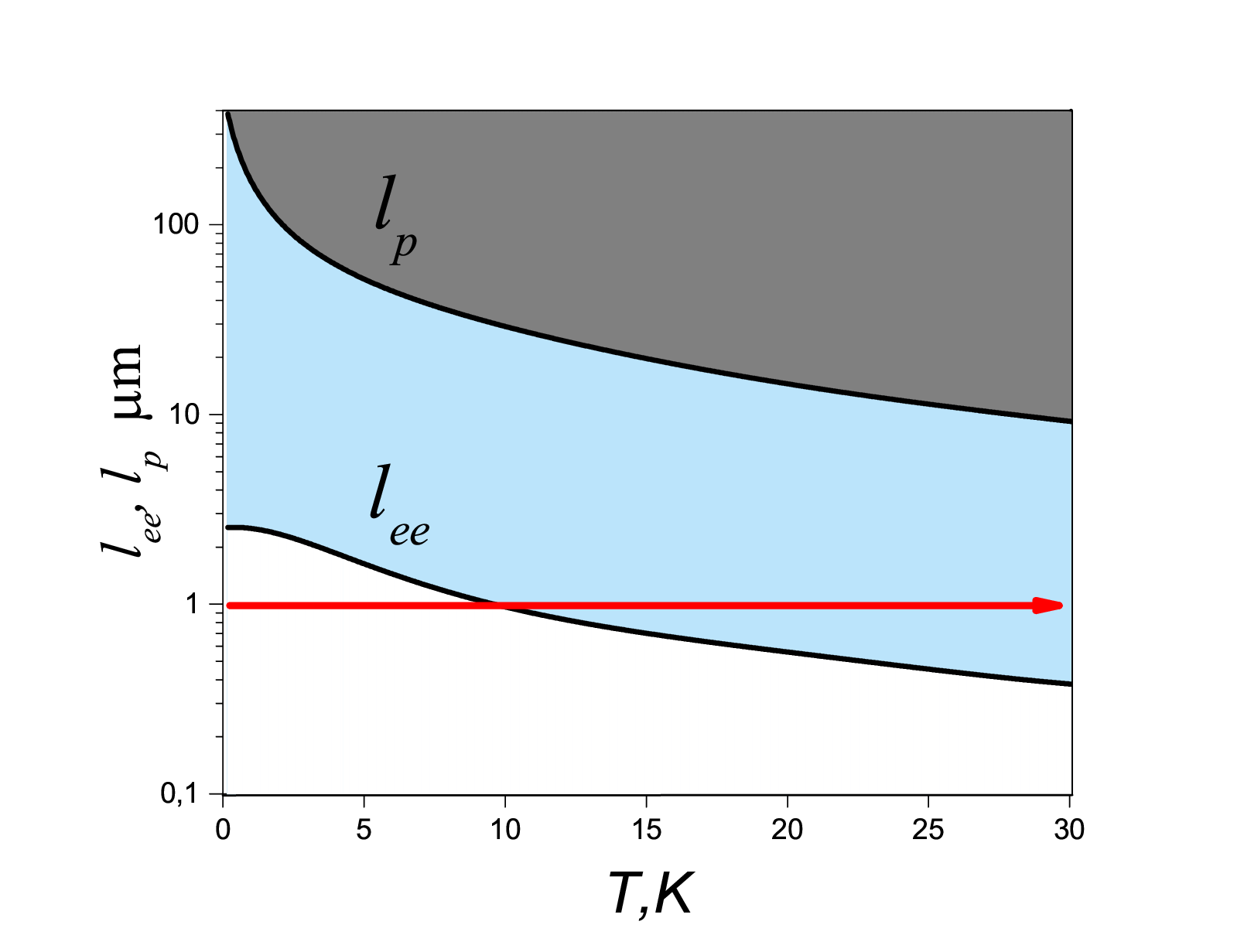} \caption[]{\label{Fig3} The scattering lengths $l_{ee}(T),l_{p}(T)$ from Ref.\cite{Cheremisin24} extended up to ambient temperatures. Arrow corresponds to fixed sample size $W=1\mu m$.}
\end{center}
\end{figure}

Let us consider the electron flow in high quality 2D system. We refer to study\cite{Cheremisin24} of scattering lengths $l_{p}(T),l_{ee}(T)$ evaluated for high mobility $\mu=2.3\cdot 10^{7}$cm$^{2}$/Vs and density $n=2.5 \cdot 10^{11}$cm$^{-2}$ 2DES hosted in a GaAs/AlGaAs quantum well. The scattering lengths are shown in Fig.\ref{Fig3} for restricted temperature range $T<30K$. Let us assume a fixed size $W=1\mu$m of the sample. Upon temperature enhancement depicted schematically by an arrow the ratios $l_{p}/W$ and $l_{ee}/W$ change continuously, hence give a certain trajectory in the diagram seen in Fig.\ref{Fig2}. According to Figs.\ref{Fig2},\ref{Fig3}, the transport regime changes sequentially, starting from ballistic to viscous and, then to combined mode. Our special interest concerns viscous flow regime shown by blue in both plots. For suggested sample size viscous transport mode starts at $T\sim 10K$.

Highly inspired by Helmholtz's finding we intend to implement his idea for 2D electron gas. In Fig.\ref{Fig1}b we present 2D analog of Helmholtz's resonator built on inverse\cite{Shur89} AlGaAs/GaAs heterostructure. The band diagram is depicted in Fig.\ref{Fig1},c, where $\delta$ denotes the quantum well thickness. Referring to acoustic analog, we draw a bottle shaped mesa which defines the contour of quantum well with 2D electron gas embedded. The metallic gate is deposited on GaAs backside. The gate to quantum well separation $d$ is assumed to be less than in-plane size of the sample. The neck part is connected to the voltage source which, in turn, is attached to the backside gate. The applied voltage is used to control 2D carriers density. Importantly, a small alternative current $i_{\sim}$ can pass through the resonator neck. This current would be considered as output signal of the resonator. Vice versa, the resonator can be excited by ac signal induced, for example, by external source of electromagnetic radiation.

We argue that the frequency of 2D Helmholtz resonator can be easily found. Indeed, using in Eq.(\ref{3H_Helmholtz}) the velocity replacement $c\rightarrow v$ and, then taking into account the narrow well condition $\delta \rightarrow 0$, the Helmholtz resonator frequency yields
\begin{equation}
\omega_{2H} =v\sqrt{\frac{w}{S l}},
\label{2H_Helmholtz}\\
\end{equation}
where $S=WL$ is the cavity area, $w$ is the neck width. For rectangular neck $w \simeq l$ Eq.(\ref{2H_Helmholtz}) provides wavelength $\lambda \sim 2\pi \sqrt{S}$ which is greater than the typical in-plane size. Noticeably, the resonator reveals ultra-high quality factor $Q=\lambda/\delta \gg 1$. Referring to acoustic analog we remind that the 2D cavity shape is arbitrary while the neck is \emph{optional}.
\section{Two-dimensional THz whistle}
\label{Two-dimensional THz whistle}
\begin{figure}[tbp]
\begin{center}\leavevmode
\includegraphics[width=1.0\linewidth]{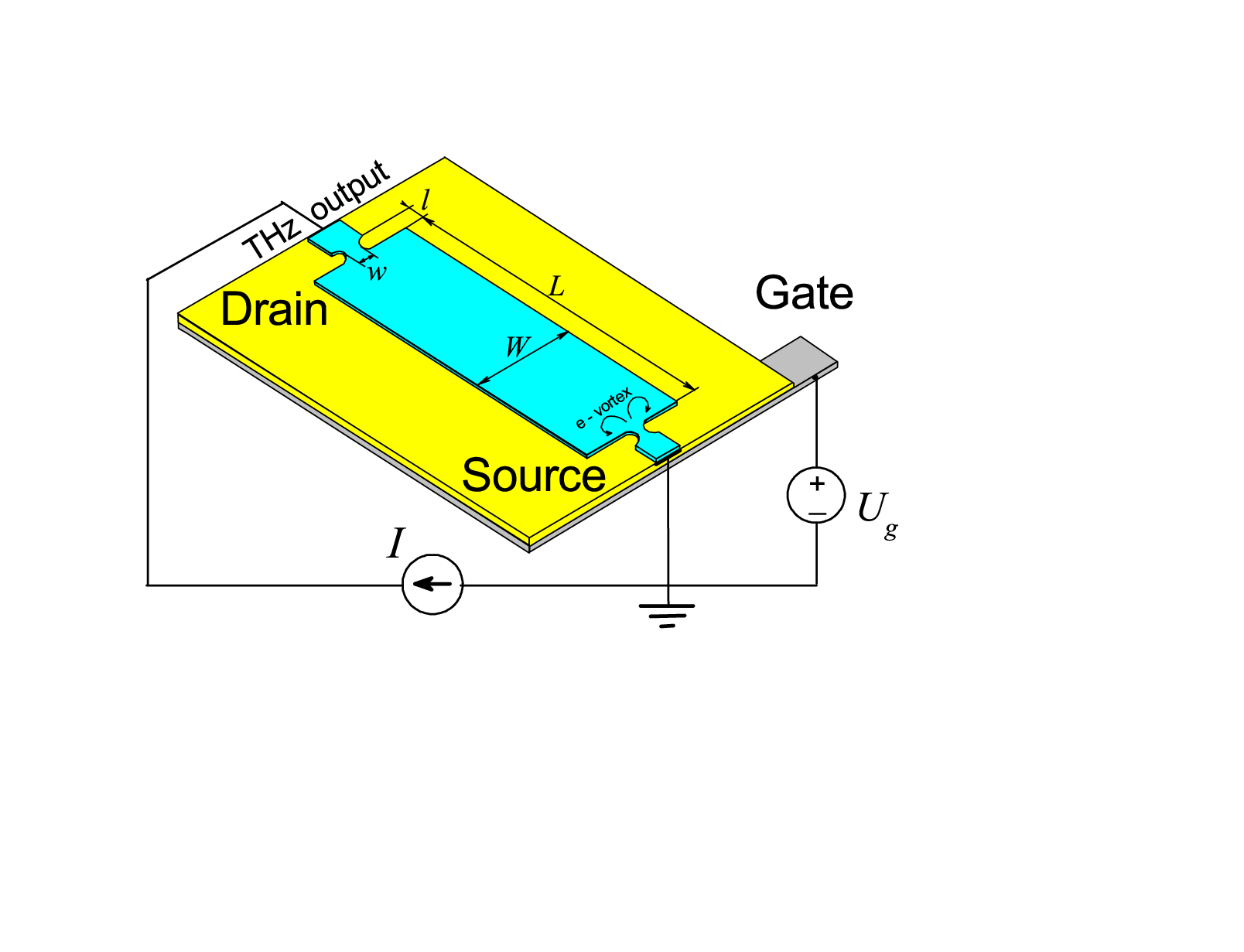} \caption[]{\label{Fig4} a. The construction and electrical assembly of THz-generator.}
\end{center}
\end{figure}

It appears that two-dimensional Helmholtz resonator is feasible, therefore its excitation would not be difficult. Recall that human whistle is governed by the jet instability under vortex formation. One may expect the same phenomena for 2D electron fluid. Therefore, we propose the turnable source of electromagnetic radiation called further THz whistle. The device is shown in Fig.\ref{Fig4}. It consists of a stripe mesa with input(output) smooth-edged orifices at the source(drain) terminal. The carrier flux between the source and drain is provided by current source.

Similar to human whistle\cite{Wilson1970} the electron flux at the source terminal, if exceeds a certain threshold, may become unstable under vortex formation. The cavity acts as Helmholtz resonator while the orifice at the drain terminal serves as emitter of output electromagnetic radiation. Referring to acoustic experiments\cite{Wilson1970} the input(output) terminals could be manufactured identical while the output frequency follows exactly Eq.(\ref{2H_Helmholtz}).

Let us estimate the important parameters of the proposed THz whistle. For simplicity, we will assume the rectangular cavity $W,L=1\mu$m and identical orifices at the source and drain $w,l=0.1\mu$m. To provide perfect screening we assume the lowest reported value\cite{Kukushkin15} of gate-to-2D channel spacing $d=50$nm. For gate voltage excess $U_{g}-U_{T}=1$V, the electron effective mass $m=0.067m_{0}$ and GaAs permittivity $\varepsilon=12.8$ we estimate the carrier density $n=1.4\cdot 10^{12}$cm$^{-2}$, plasma wave velocity $v=1.62\cdot 10^{8}$cm/s and, finally, the output frequency $f=\frac{v}{2\pi W}=0.26$THz. According to Fig.\ref{Fig3} the e-e scattering length is less than all other length scales at $T\geq 10$K justifying viscous transport regime. Importantly, an increase in temperature simplifies our ambitious task. Indeed, the diagram in Fig.\ref{Fig3} demonstrated that viscous transport mode $l_{p}\gg W\gg l_{ee}$ may persist even for elevated temperatures. The enhanced current, if applied, may lead to vortex formation and, in turn, whistling in THz range. Referring to flux velocity limited by saturation value $V\sim 10^{7}$cm/s in GaAs we estimate atop the Mach number as $M=V/v=0.06$. Surprisingly, this estimate is close to range $M=0.02-0.04$ reported in acoustic experiment\cite{Wilson1970}. Also, the Strouhal number ${\text S}=\frac{V}{lf}=3.8$ is close to that $\sim 0.6$ reported for acoustic whistling\cite{Wilson1970}. Finally, assuming much higher temperatures $T\gg 30$K we estimate the electron viscosity $\nu\sim \hbar/m =12$cm$^{2}$/s caused by interelectronic collisions, i.e. when $l_{ee}\sim n^{-1/2}$. The respective Reynolds number ${\text R}=\frac{V l}{\nu}=8$ confirms the possibility of nonlinear hydrodynamic effects, namely the desired vortices formation.

\section{Conclusions}
\label{Conclusions}
In conclusion, we propose the turnable THz generator built on two-dimensional analog of conventional Helmholtz resonator. The excitation of THz-generator
is provided by the current flow similar to whistle in conventional hydrodynamics. The output frequency, hydrodynamic parameters of 2D THz-whistle are discussed.

\bibliography{THZ_HELMHOLTZ}

\end{document}